\documentclass[jkps,preprint,fleqn,showpacs,showkeys]{revtex4}
\usepackage{graphicx}
\usepackage{amssymb}
\usepackage{amsmath}
\usepackage{bm}
\begin{document}
\setcounter{page}{0}
\title[]{Classical Mechanics of Collinear Positron-Hydrogen Scattering}
\author{Min-Ho \surname{Lee}}
\author{Chang Woo \surname{Byun}}
\author{Jin-Sung \surname{Moon} }
\author{Nark Nyul \surname{Choi}}
\email{nnchoi2001@gmail.com}
\affiliation{School of Liberal Arts and Teacher Training, Kumoh National Institute of Technology, Gumi 730-701, Korea}
\author{Dae-Soung \surname{Kim}}
\affiliation{Department of Global Education, Gyeonggi College of Science and Technology, Siheung 429-792, Korea}
%
\date[]{7 April 2015}
\begin{abstract}
We study the classical dynamics of the collinear positron-hydrogen scattering system below the three-body breakup threshold. Observing the chaotic behavior of scattering time signals, we introduce a code system appropriate to a coarse grained description of the dynamics. And, for the purpose of systematic analysis of the phase space structure, a surface of section is introduced being chosen to match the code system. Partition of the surface of section leads us to a surprising conjecture that the topological structure of the phase space of the system is invariant under exchange of the dynamical variables of proton with those of positron. It is also found that there is a finite set of forbidden patterns of symbol sequences.  And the shortest periodic orbit is found to be stable, around which invariant tori form an island of stability in the chaotic sea. Finally we discuss a possible quantum manifestation of the classical phase space structure relevant to resonances in scattering cross sections.
\end{abstract}
\maketitle
\section{INTRODUCTION}
The three-body problem of Coulombic systems is one of the most fundamental problem of theoretical physics. A large number of papers has been published which deals with the classical mechanics as well as the quantum mechanics of electron-hydrogenic atom systems, that is two-electron atoms \cite{Wannier1953, Gregor2000, Lee2010}. The classical motion of three bodies in a two-electron atom above the three-body breakup threshold is known to be regular although it is nonintegrable \cite{McGehee1974}, and the behavior of electron-impact ionization of hydrogenic atoms near the breakup threshold has been treated long time \cite{Rost1998, Friedrich2006}. In contrast to the dynamics above the threshold, the classical motion below the threshold is no longer regular; the classical phase space consists of not only reguar but also chaotic parts \cite{Richter1993}. The chaotic dynamics below the threshold produces infinitely many resonances in photoelectron spectra of two-electron atoms, and the fluctuating parts of the spectra owing to the resonances show a series of seemingly meaningless irregular signals \cite{Jiang2008}. However, by the help of modern techniques of semiclassical mechanics, the fluctuating parts can be characterized by the actions and stabilities of the infinite number of closed triple collision orbits (CTCO) \cite{Byun2007, Lee2010}.

Positron-hydrogen scattering system is another three-body Coulomb problem, which has recently attracted a growing interest as powerful positron beams have developed since the discovery of low energy positron beams \cite{Cherry1958, Charlton2001, Surko2005, Bhatia2014}. The behavior positron-impact ionization of hydrogenic atoms above the breakup threshold, whose corresponding classical dynamics is regular as in the electron impact ionization, has been examined by both classical and quantum mechanical methods \cite{Rost1994, Ihra1997, Kadyrov2007, Jansen2009}. Considering there are a large amount of papers for the classical dynamics of two-electron atoms for wide range of energies including below three-body breakup threshold, however, it is noteworthy that only a limited number of papers deal with classical dynamics of positron scattering by a hydrogen target below the threshold, where the classical motion might be no longer regular and thus fluctuations in quantum scattering cross sections might occur as in the case of two-electron atoms.

Stimulated by a prominent role of classical calculations in success of semiclassical interpretations and predictions on quantum mechanical properties of two-electron atoms \cite{Richter1993}, a study was conducted of classical dynamics of positron-hydrogen scattering for wide range of energies including below as well as above the threshold \cite{Spivack1999}. However, in that work, the classical mechanics was merely used as a calculational method for Monte Carlo simulations rather than being analyzed for the purpose of characterization of the resonance structures as was done in two-electron atoms. A systematic analysis of the classical phase space of the positron-hydrogen scattering system has not been given yet for the energies relevant to the resonances, that is for energies below the breakup threshold \cite{Surko2005, Varga2008}.

Resonances lying close to the threshold are assumed to be associated with CTCOs as demonstrated in papers deals with photoelectron spectra of two-electron atoms \cite{Byun2007, Lee2010, Tanner2007}. By virtue of the fact that CTCOs are completely included in the collinear invariant subspace, study of classical dynamics in the subspace would give a great insight into the resonances in positron-hydrogen scattering in the real three-dimensional world.

In this paper we will report on our investigation of the classical dynamics of positron scattering by a hydrogen target in the collinear $p^+e^-e^+$ configuration for energies below the three-body breakup threshold. The paper is organized as follows. In Sec. II the Hamiltonian is introduced, and a transformation is also introduced to regularize Coulomb singularity at the binary collisions for proper numerical treatment of the classical equations of motion. In Sec. III scattering time signals are given as a function of the initial state of the hydrogen atom at the instant the positron starts coming to the hydrogen atom at a fixed distance. Based on observation of the dips in the scattering time signal, a code system is introduced for classifying orbits, and a surface of section (SOS) is also introduced for analyzing the phase space structure. In Sec IV we present our findings based on the calculational result of the partition of SOS showing the phase space structure. Discussions are also given in that section. Finally, we conclude with a summary in Sec. V.

\section{Equations of Motion}
We consider the collinear $p^+e^-e^+$ configuration in which the proton and the positron are located on the opposite side of the electron. We further assume that the proton is infinitely heavy and fixed at the origin. In atomic units, this system is described by the Hamiltonian
\begin{align}
   H &=\frac{1}{2} p_1^2 - p_1 p_2 + p_2^2  - \frac{1}{x_1} - \frac{1}{x_2} + \frac{1}{x_1+x_2}
\end{align}

where $x_1$ and $x_2$ are respectively the distances of the proton and the positron from the electron, and their conjuate momenta are denoted by $p_1$ and $p_2$.

For proper numerical treatment, the singularities at $x_1 =0$ or $x_2=0$ are regularized by introducing a fictitious time $\tau$ and a transformed Hamiltonian $K$ such as \cite{Arseth1974}
\begin{align}
   dt &= x_1 x_2 d\tau \\
   K &= x_1 x_2 ( H - E )
\end{align}

In addition, introducing regularized coordinates $Q_1$, $Q_2$, and momenta $P_1$, $P_2$ such that
\begin{equation}
   x_1 = Q_1^2, \quad
   x_2 = Q_2^2, \quad
   p_1 = \frac{P_1}{2Q_1}, \quad
   p_2 = \frac{P_2}{2Q_2}
\end{equation}
one can obtain the equations of motion as follows
\begin{align}
  \dot{Q}_1 &= \frac{1}{4} P_1 Q_2^2 - \frac{1}{4} Q_1 Q_2 P_2 \\
  \dot{Q}_2 &= -\frac{1}{4} Q_1 Q_2 P_1 + \frac{1}{2} P_2 Q_1^2 \\
  \dot{P}_1 &= \frac{1}{4} Q_2 P_1 P_2 + 2 Q_1  - \frac{1}{2} P_2^2 Q_1 + 2 Q_1 Q_2^2 \left [ E - \frac{Q_2^2}{(Q_1^2 + Q_2^2)^2} \right ]  \\
  \dot{P}_2 &= \frac{1}{4} P_1 Q_2 (Q_1 P_2 - P_1) + 2 Q_2 + 2Q_1^2 Q_2 \left [ E -  \frac{Q_1^2}{(Q_1^2 + Q_2^2)^2} \right ]
\end{align}
where the dot notation is used for the fictitious-time derivative.

\section{Symbolic Code System and Surface of Section (SOS)}
\begin{figure}
\includegraphics[width=\textwidth]{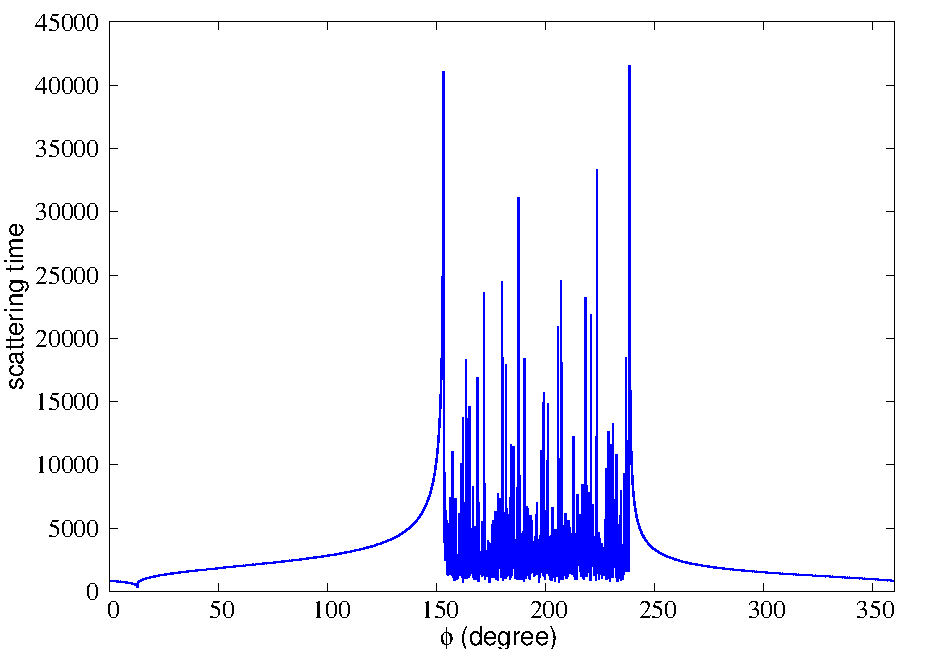}
\caption{(Color Online) The scattering time as a function of the angle variable ($\phi$) for the total energy $E=-1$ and the energy of the hydrogen atom $E_1=-1.2$.
}\label{fig1}
\end{figure}

It has been found useful to examine scattering time signals as a starting step to systematic investigation of the classical dyamics of three-body problems \cite{Choi2004, Lee2005}. In Fig. \ref{fig1}, a typical signal for scattering time is shown as a function of the angle variable ($\phi$) of the action-angle variable pair of the one-dimensional hydrogen atom at time $t=0$. The energy of the hydrogen atom was chosen to be $E_1 = -1.2$, and the total energy of the scattering system to be $E=-1$. The positron started inward at the distance of 100 from the proton, i.e. at $x_1 + x_2 = 100$ and we computed the trajectories until the positron reached $x_1 + x_2 = 1000$ after scattering. Note that the scattering time is not the fictitious time but the real time spent by the trajectory.

The scattering time signasl show typical chaotic scattering signals with infinitely many peaks and dips, and we focus on the dips as we did in two-electron atoms\cite{Lee2005}. Note that two wings stretch from the primary dip located at $\phi=\phi_d \approx 12.61$. as can be seen in Fig. \ref{fig1}. The left wing stretches beyond $0^\circ$, and its end reaches a boundary (at $\phi \approx 240^\circ$) of the region of chaotic signals; the right wing stretches to the other bounday at $\phi \approx 150^\circ$. The wings rise infinitely at the ends. And, by examining many other dips in the region of chaotic signals, we confirmed that every dip has its own two wings, and every wing rises infinitely at its end.

\begin{figure}
\includegraphics[width=0.46\textwidth]{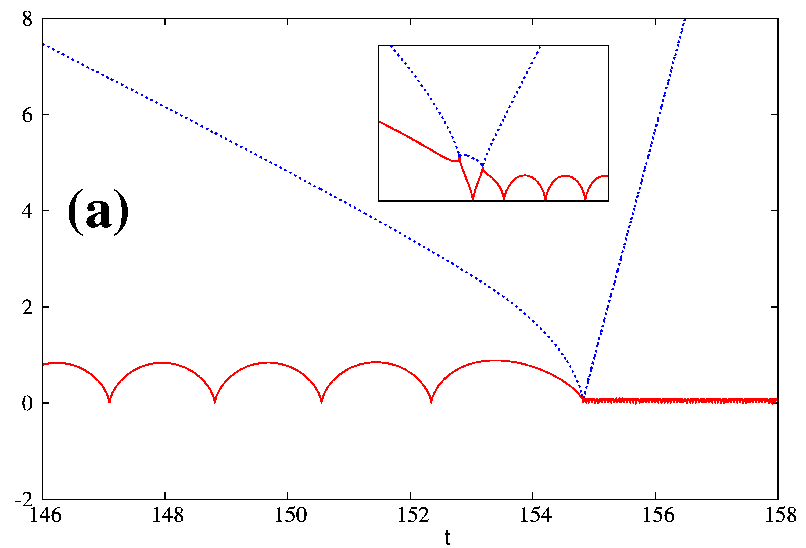}
\includegraphics[width=0.4\textwidth]{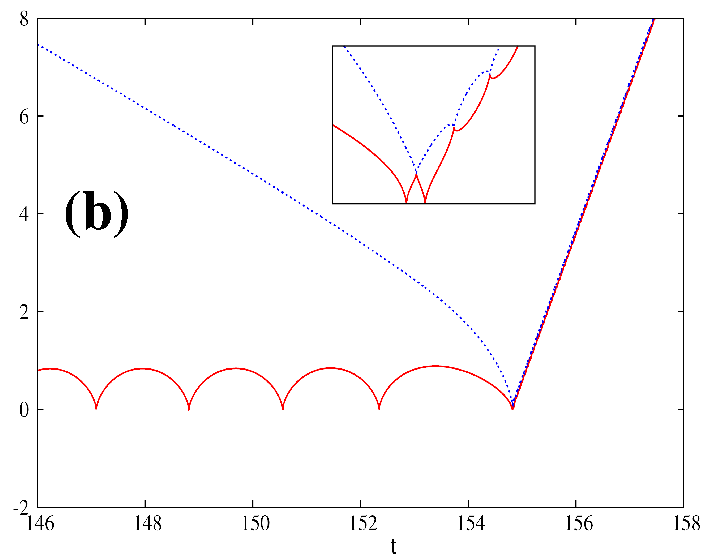}
\caption{(Color Online) Scattering trajectories evolved with the initial conditions (a) $\phi=12.58 < \phi_d$ and (b) $\phi=12.64 > \phi_d$. Solid (red) line for electron's position $x_1$ as a function of time, and dotted (blue) line for positron's position $x_1 + x_2$. Insets are magnifications around the instant when trajectories nearly approach the triple collision.
}\label{fig2}
\end{figure}

 In Fig. \ref{fig2}, two scattering trajectories are depicted with the initial conditions slight deviating from the primary dip; (a) one with the initial condition $\phi < \phi_d$, and (b) the other with $\phi > \phi_d$. From these plots, we can see that the trajectory with the initial condition corresponding to a dip would undergo the triple collision and furthermore the triple collision separates two qualitatively different scattering products, i.e. inelastic and exchange scattering as in Figs. \ref{fig2}(a) and (b) respectively. By examining trajectories with initial conditions around many other dips in the region of chaotic signals, we confirmed that every dip is produced by the trajectories ending in triple collision and every pair of wings stretching from a dip are branches to different scattering products.


Symbolic dynamics is a useful tool for a classification of orbits in nonintegrable systems \cite{Hao1998, Bai1998}. From the observation of the above-mentioned separation of dynamical behavior by the triple collision, we naturally introduce a symbol code system by defining a symbolic assignment such as
Symbolic dynamics is a useful tool for a classification of orbits in nonintegrable systems \cite{Hao1998, Bai1998}. From the observation of the above-mentioned separation of dynamical behavior by the triple collision, we naturally introduce a symbol code system by defining a symbolic assignment such as
    \begin{equation*}
\begin{cases}
    1 & \text{ if the trajectory undergoes a binary collision of the proton and electron }\\
    2 & \text{ if the trajectory undergoes a binary collision of the positron and electron} \\
    \text{c} & \text{ if the trajectory undergoes the triple collision}
\end{cases}
\end{equation*}
For example, the scattering orbit in Fig. \ref{fig2}(a) can be represented by a symbol sequece ...111212111... while the orbit in Fig. \ref{fig2}(b) by ...11121222... It is noticeable  that the triple collision is a non-regularizable singularity and thus we cannot evolve trajectories any more after undergoing the triple collision \cite{McGehee1974, Richter1993}. Thus we let triple collision orbits terminate with the symbol c.

\begin{figure}
\includegraphics[width=\textwidth]{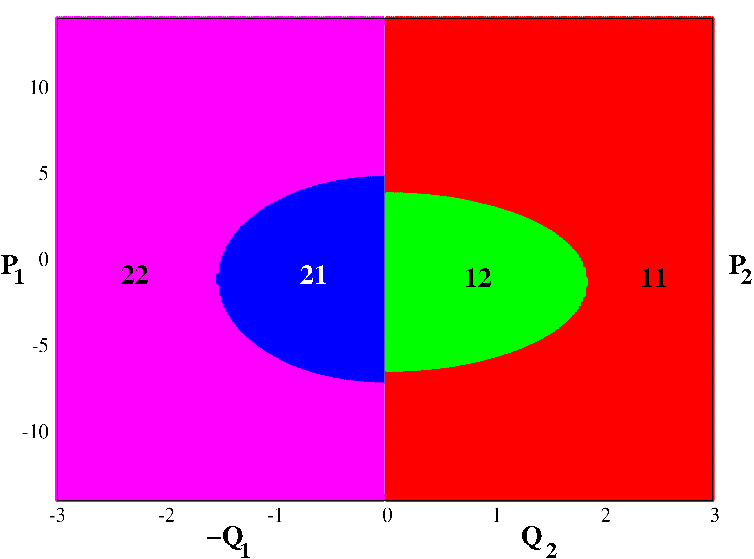}
\caption{(Color Online) Partition of SOS induced from classification of orbits by symbol sequences of length 2. The SOS consists of two sheets, $x_1=0$ (the right half-plane) and $x_2=0$ (the left half plane). Points represented by different symbol sequences are painted with different colors. Variables $Q$'s and $P$'s are the regularized coordinates inroduced in Sec. II.
}\label{fig3}
\end{figure}

It is convenient to explore the phase space structure by using the Poincare section method \cite{Richter1993, Bai1998, Lee2005b}. It is noted that, by virtue of the energy scaling property of Coulomb systems, we only need to deal with the system for a fixed total energy $E=-1$ \cite{Gregor2000}. Based on the obvious fact that all orbits except the shortest CTCO undergo binary collisions ($x_1 =0$ or $x_2 = 0$) at least once, the set {$x_1 =0$ or $x_2 = 0$} composed of two sheets of binary collisions is chosen as the SOS (see Fig. \ref{fig3}). The Poincare mapping is then naturally defined as $F(X) = Y$ if a trajectory starts at $X \in$ SOS consecutively crosses at $Y \in$ SOS.

A partition of SOS can be induced by combining the Poincare mapping $F$ and the symbol code system introduced above. For an orbit starting from $X \in$ SOS, a symbolic sequence of finite or infinite length can be given to it as
\[
  S(X) = \sigma_0 \sigma_1 \dots \sigma_k \dots
\]
where
\begin{equation*}
\sigma_k = \begin{cases}
             1 & \text{if $F^k(X)$ lies in the sheet $x_1=0$} \\
             2 & \text{if $F^k(X)$ lies in the sheet $x_2=0$} \\
             c & \parbox[t]{.6\textwidth}{ if the orbit starting from $ F^{k-1}(X)$ falls in triple collision without any more crossing of SOS.}
           \end{cases}
\end{equation*}

In Fig. \ref{fig3}, we present a coarse-grained partition of SOS, where the whole SOS is partitioned to four regions of the sequences of length 2. The borders of the regions are the stable manifold of the triple collision which consists of orbits terminating with the symbol c. In the following section, we will present further finer partitions of SOS, which make it possible to find remarkable dynamical properties of the system.

\section{Results and Discussion}
\begin{figure}
\includegraphics[width=0.46\textwidth]{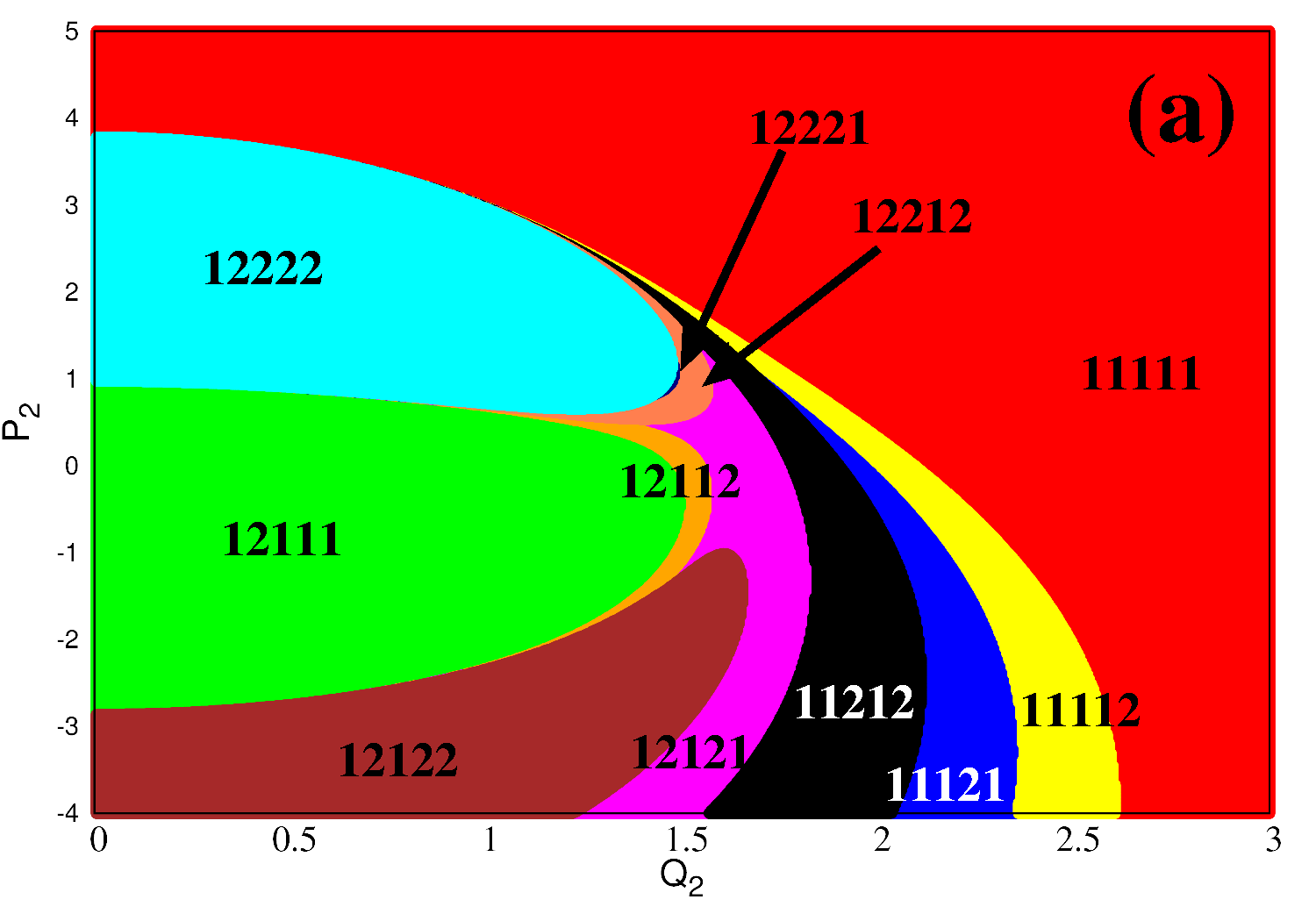}
\includegraphics[width=0.46\textwidth]{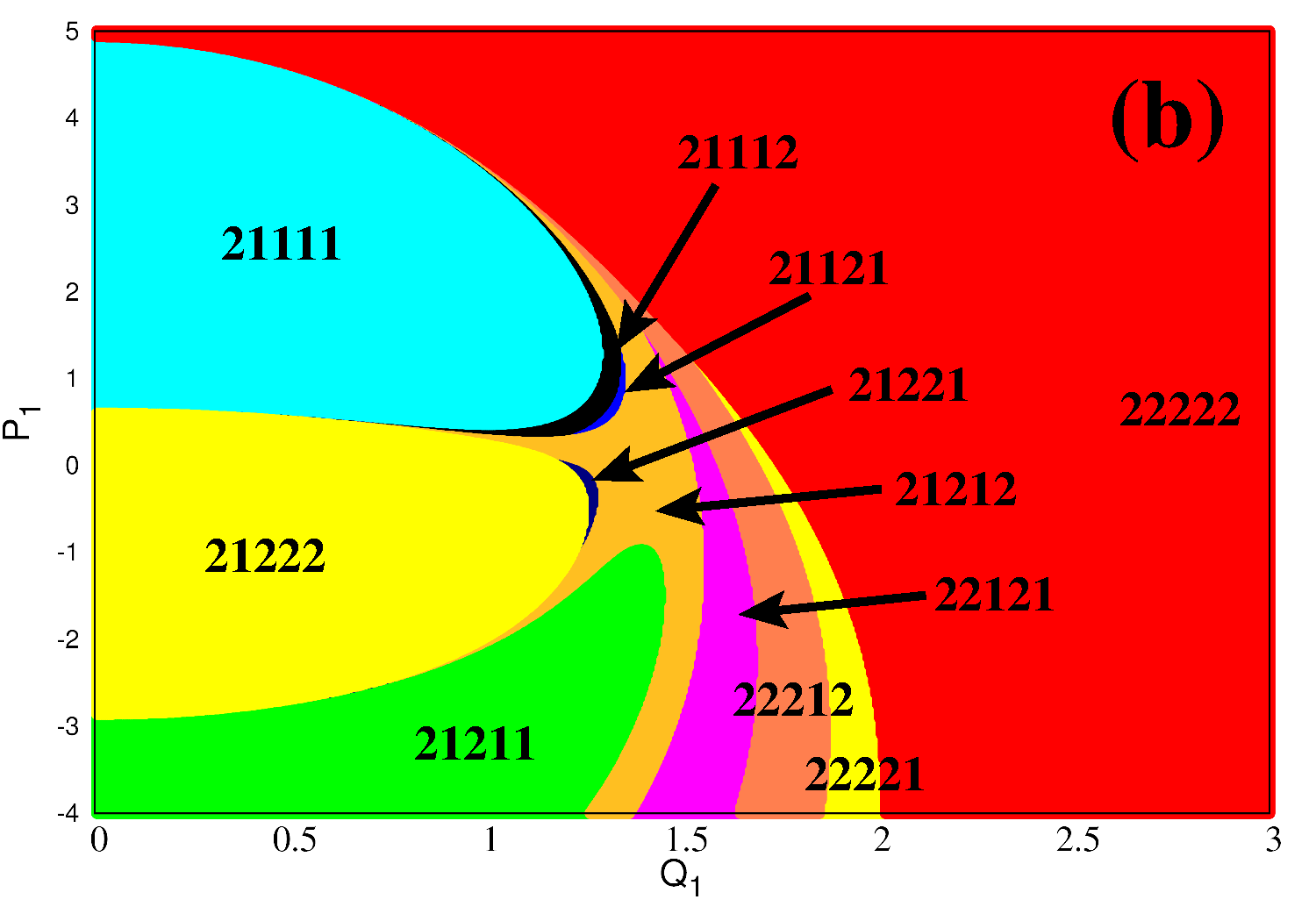}
\caption{(Color Online) Partition of SOS by symbol sequences of length 5 repsented in two sheets (a) $x_1=0$ and (b) $x_2=0$. Colors are used for the same purpose as in Fig. \ref{fig3}.
}\label{fig4}
\end{figure}

Proceeding by three step from the Fig. \ref{fig3}, we obtain a finer partition of SOS in which each region is classified by a sequence of length 5 as shown in Fig. \ref{fig4}. Comparing the two sheets (a) $x_1 = 0$ and  (b) $x_2 = 0$, we can find that they have the same geometric structure. Actually we confirmed this symmetry of geometric structure by examining the partition classified by symbolic sequences of length 7, a plot of which is not given here for brevity.  Thus we propose a conjecture that the geometric structure of the dynamics is invariant under exchange of the variables $(x_1, p_1)$ with $(x_2, p_2)$. This is a very surprising finding since the masses of proton and positron are quite different and thus equations of motion (5)-(7) is not invariant under the exchange.

Checking the symbol sequences representing the partitioned parts in Fig. \ref{fig4}, we can see that there are no trajectories representing the symbol sequences 11122, 11221, 11222, 12211, 11211, and their reflections 22211, 22112, 22111, 21122, 22122. By examining the finer partition up to symbol length 7, we conjecture that there is a finite set of forbidden patterns of symbol sequences such as {1122, 2211, 11211, 22122}. Note that there are no scattering orbits ...111222... or ...1112111...; the simplest scattering orbits are ...111212111... and ...11121222... as can be seen in Fig. \ref{fig2}.

\begin{figure}
\includegraphics[width=0.46\textwidth]{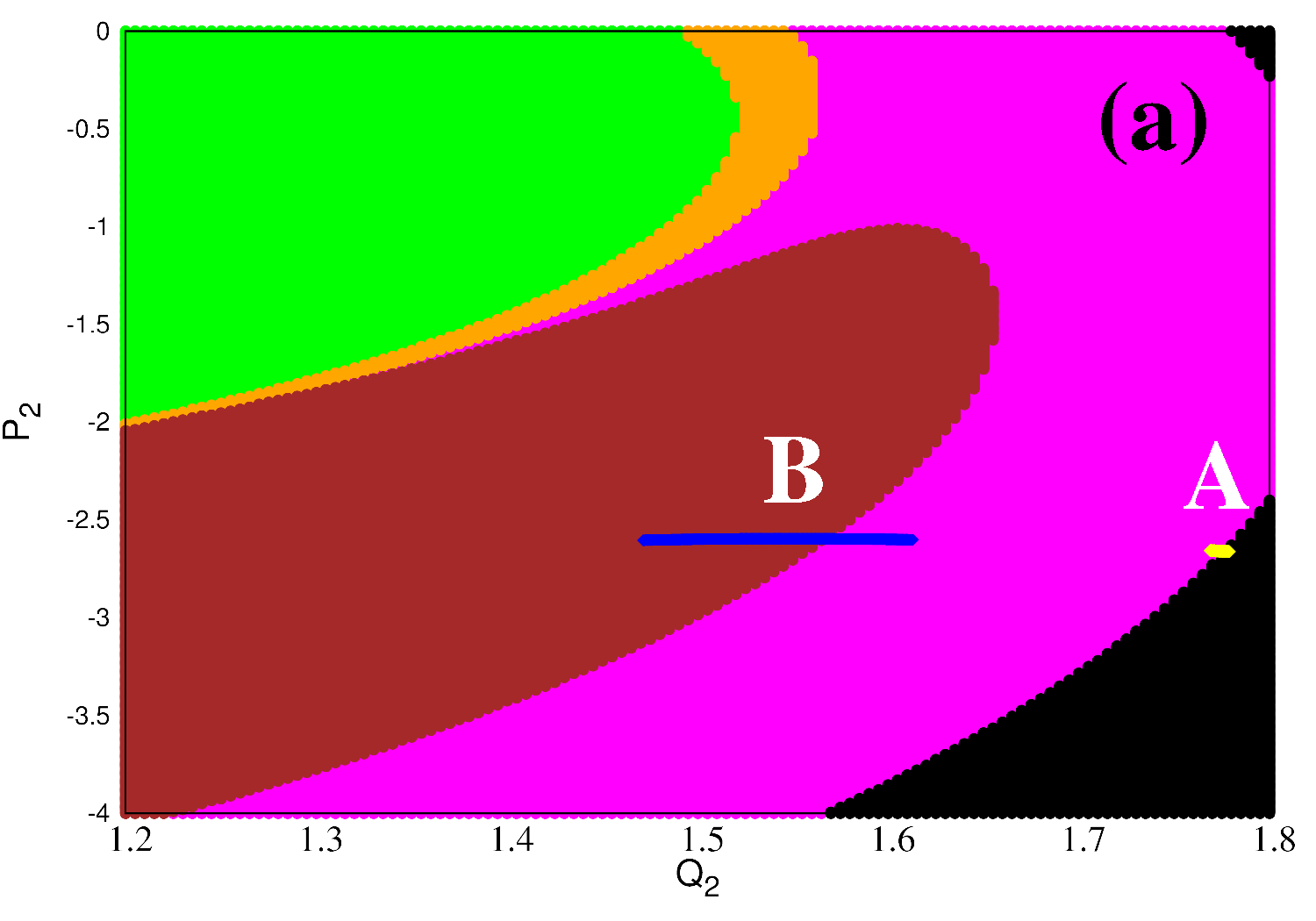}
\includegraphics[width=0.46\textwidth]{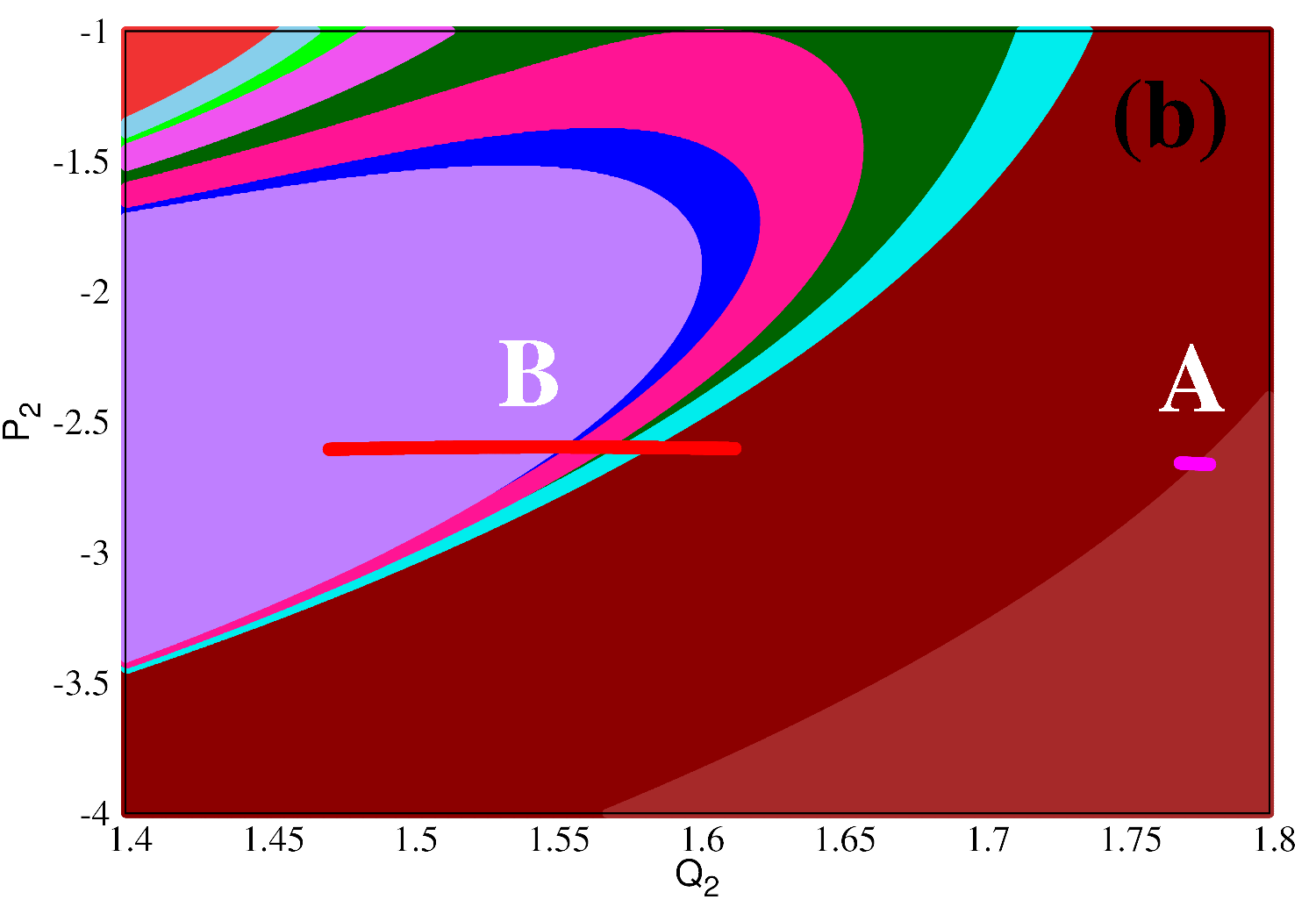}
\caption{(Color Online) Intersection of SOS and orbits with the initial conditions$\phi \in [5^\circ:15^\circ]$ and $\phi \in [150^\circ:240^\circ]$. The former is denoted by A; the latter by B. The segments A and B are plotted in the SOS partitioned by symbol sequences of (a) length 5 and (b) length 7.
}\label{fig5}
\end{figure}

The partition of SOS is also useful for understanding the main features of the scattering time signals in Fig. \ref{fig1}. Initial conditions evolve with time and pass through the SOS. A set of initial conditions at their first passage through the SOS forms an initial line segment in the SOS.
And then the initial line segment proceeds in SOS by the Poincare map. In Fig. \ref{fig5}, we show images of two initial line segments obtained by applying the Poincare map several times.
The segment A is the image of initial line segment formed from the interval $\phi \in [5^\circ:15^\circ]$ which includes the initial condition $\phi_d$ corresponding to the dip with the widest wings in Fig. \ref{fig1}; the other (denoted by B) for the interval $\phi \in [150^\circ:240^\circ]$ enclosing the initial conditions corresponding to the chaotic signals consisting of many peaks and dips in Fig. \ref{fig1}. It can be seen from Fig. \ref{fig5}(a) that both the the line segments A and B are cut by the stable manifold of tripl collision. However, looking into the partition of SOS by symbol sequences of length 7, we can find the number of intersection points of the segment B and the stable manifold of triple collision increases as the partition gets finer as in Fig. \ref{fig5}(b). This is consistent with the feature of the scattering time signals such that there are infinitely many dips in in the interval $\phi \in [150^\circ:240^\circ]$.

From studies of three-body problems, it has been generally known that trajectories in captured state can exit to uncaptured state along only the orbits falling in triple collision; the closer the trajectory gets to triple collision, the larger the escaping velocity of the trajectory gets \cite{McGehee1974, Lee2005}. In other words, the farther the trajectory gets from triple collision, the smaller the escaping velocity gets. This means that orbits with initial conditions far from dips, i.e. very close to the ends of wings escape the three-body interaction zone with zero energy of relative motion between the hydrogen atom (positronium) and electron (proton). Thus  the ends of the wings rise infinitely, producing peaks in scattering time signals as was mentioned in Sec. III.

In addition to the role of triple collision as an exit from captured state, the other prominent role of triple collision is led by time-reversal symmetry of the present system as follow: scattering trajectories can enter into captured state along only the orbits starting from triple collision \cite{Choi2004, Lee2005}. Thus, in semiclassical physics, resonances in scattering cross sections can be described in terms of contributions from CTCOs, i.e. closed orbits starting and ending in the triple collision \cite{Lee2010, Byun2007, Tanner2007, Lee2010JKPS}. It can be easily seen that all CTCOs lies in the collinear configuration, and thus the present study of classical mechanics of collinear $p^+e^-e^+$ system is relevant to the fluctuations in the scattering cross sections in the real 3-dimensional space. Therefore, as was done for two-electron atoms \cite{Lee2010, Byun2007}, significant predictions can be made on the total scattering cross section of positron by the hydrogen atom with the total energies approaching the three-body breakup threshold from below as follows: (1) fluctuation in the cross section as a function of the total energy would be chaotic, (2) the amplitude of the fluctuation decays algebraically as the energy approaches the three-body breakup threshold, and (3) the Fourier transform of the fluctuation would show distinct peaks at the position of the actions of CTCOs.

The statement (1) comes from the chaotic nature of the symbolic dynamics of the system: between any two ordinary admissible symbol sequences, we can find arbitrarily  long symbol sequences with ending in triple collision.
The statement (2) comes from the fact that the instabilities of relevant part of CTCOs approach to infinity as the energy approaches the breakup threshold, which is related to the irregularizability of the triple collision \cite{Lee2010, Byun2007}. The exponent for algebraic decay is determined by the stability of the relevant part of CTCOs, which will be obtained in future work. Finally, the statement (3) is based on closed orbit theory \cite{Du1988}, which is a general theory for atomic photoabsorption cross sections. However, the condition required in the theory is nothing but the restriction on the starting and ending point of the leading orbits to captured state: both the starting and ending point should be close to the center of atomic system. Thus it can be applied to scattering cross sections for the systems with CTCOs. The action values and relative instabilities of CTCOs are not calculated in the present paper, remaining as a future work. However, we can predicted that there are no peaks in Fourier transform of the fluctuation which correspond to the forbidden symbol sequences c112c, c122c, c221c, c211c, etc. This is quite important characteristics different from two-electron atoms, reflecting the phase space structure of the present system.

\begin{figure}
\includegraphics[width=\textwidth]{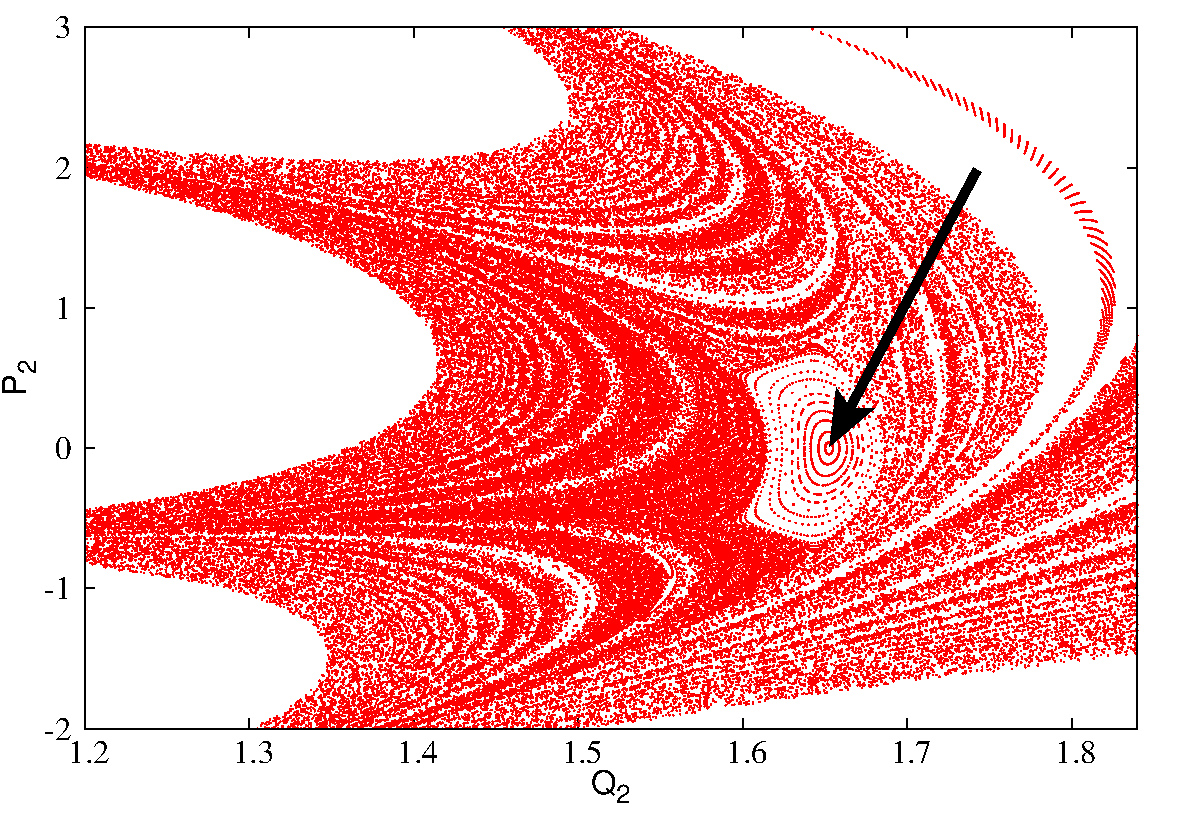}
\caption{A phase portrait revealing an island of stability in the chaotic sea. The center of the tori is indicated by an arrow.
}\label{fig6}
\end{figure}
Looking into the partition of SOS, we get doubt if there are some islands around the point $(Q_2, P_2) \approx (1.64, 0.0)$ in Fig. \ref{fig4}. In order to lift obscurity, we draw a phase portrait around that point. As can be seen in Fig. \ref{fig6}, there is actually an island of stability consisting of tori in the chaotic sea. At the center of the island which is indicated by an arrow in Fig. \ref{fig6}, there is a stable periodic orbit, which is found to be the shortest one with infinite symbol sequence $\overline{12}$; all orbits in the island is represented by the same symbol sequence. This is another characteristics different from two-electron atoms of which collinear eZe subspace is known to be fully chaotic \cite{Richter1993, Bai1998}. Since it has been accepted that there is no bound state of positron-hydrogen system in the real 3-dimensional world \cite{Bhatia2014}, the stability of the tori in the direction orthogonal to the collinear subspace should be investigated for semiclassical prediction of existence or absence of bound or long-lived states.
And it will be also a good subject of future study to understand effects of the tori through dynamical tunneing in the scattering cross sections \cite{Tomsovic1994}.

For complete study of positron-hydrogen system in future, we should mention on a scattering channel which is excluded in the present work, i.e. positron-electron annihilation. It is well known that the possible contribution of the positron-electron annihilation to the total scattering cross section is negligible except in the limit of zero positron energy \cite{Charlton2001}. However,it may be necessary to consider positron-electron annihilation for total energies just below the three-body breakup threshold since the amplitude of fluctuating part in the total cross section would decay to zero as the total energy approaches the breakup threhold and thus would be affected even by small perturbations owing to the annihilation channel.

\section{Conclusion}
The classical dynamics of positron scattering by the hydrogen atom in the collinear configurational subspace was studied for total energies below the three-body breakup threshold. A systematic analysis of the phase space structure of the system was given for the first time, resulting in important new findings: (1) the system shows chaotic scattering, (2) the topological structure of the phase space is invariant under exchange of the dynamical variables of proton with those of positron, (3) there is a finite set of forbidden patterns of symbol sequences such as {1122, 2211, 11211, 22122}, (4) there are tori forming a stable island around the shortest periodic orbit. And, based on the phase space structure and well-known general dynamics near triple ollision.we explained main features of scattering time signals.

Using the relevance of CTCOs to fluctuations in scattering cross sections  owing to resonances, quantum manifestations of the classical dynamics were predicted on the total scattering cross section such that (1) fluctuation in the cross section as a function of the total energy would be chaotic, (2) the amplitude of the fluctuation decays algebraically as the energy approaches the three-body breakup threshold, and (3) the Fourier transform of the fluctuation would reflect the phase space structure through absence of peaks represented by the symbol sequences c112c, c122c, c221c, c211c, etc. including the forbidden symbol sequences.

This work is considered as a first step to full understanding of classical, semiclassical and quantum dynamics of positron-hydrogen system, and we suggested several subjects to be done in future. In addition, it would be a kind of challenge to find a quantum mechanical menifestation of the structural invariance of the phase space under exchange of the dynamical variables of proton with those of positron.

\begin{acknowledgments}
This work was supported by the Kumoh National Institute of Technology under contract number 2011-104-015.
\end{acknowledgments}


\begin{references}
\bibitem{Wannier1953} G. H. Wannier, Phys. Rev. 90, 817 (1953).
\bibitem{Gregor2000} G. Tanner, K. Richter, and J. M. Rost, Rev. Mod. Phys. 72, 497 (2000).
\bibitem{Lee2010} M.-H. Lee, C. W. Byun, N. N. Choi, G. T. Tanner, Phys. Rev. A 81, 043419 (2010).
\bibitem{McGehee1974} R. McGehee, Invent. Math. 27, 191 (1974).
\bibitem{Rost1998} J.-M. Rost, Phys. Rep. 297, 271 (1998).
\bibitem{Friedrich2006} H. Friedrich, {\it Theoretical Atomic Physics} (Springer, Heidelberg 2006), Chap. 4.
\bibitem{Richter1993} K. Richter, G. Tanner, and D. Wintgen, Phys. Rev. A 48, 4182 (1993).
\bibitem{Jiang2008} Y. H. Jiang, R. P\"{u}ttner, D. Delande, M. Martins, and G. Kaindl, Phys. Rev. A 78, 021401(R). (2008).
\bibitem{Byun2007} C. W. Byun, N. N. Choi, M.-H. Lee, and G. Tanner, Phys. Rev. Lett. 98, 113001 (2007).
\bibitem{Cherry1958} W. H. Cherry, {\it Secondary electron emmision produced from surfaces by positron bomardment} (Ph.D. Thesis, Princeton University 1958).
\bibitem{Charlton2001} M. Charlton and J. W. Humberston, {\it Positron Physics} (Cambridge University Press, Cambridge 2001).
\bibitem{Surko2005} C. M. Surko, G. F. Gribakin, and S. J. Buckman, J. Phys. B 38, R57 (2005).
\bibitem{Bhatia2014} A .K. Bhatia, J. Atom. Mol. Cond. Nano Phys. 1, 45 (2014).
\bibitem{Rost1994} J.-M. Rost and E .J. Heller, Phys. Rev. A 49, R4289 (1994).
\bibitem{Ihra1997} W. Ihra, J. H. Macek, F. Mota-Furtado, and P. F. O'Mahony, Phys. Rev. Lett. 78, 4027 (1997).
\bibitem{Kadyrov2007} A. S. Kadyrov, I. Bay, and A. T. Stelbovics, Phys. Rev. Lett. 98, 263202 (2007).
\bibitem{Jansen2009} K. Jansen, S. J. Ward, J. Shertzer, J. H. Macek, Phys. Rev. A 79, 022704 (2009).
\bibitem{Spivack1999} O. R. Spivack, J. Phys. IV France 9, Pr6-191 (1999).
\bibitem{Varga2008} Phys. Rev. A 77, 044502 (2008).
\bibitem{Tanner2007} G. Tanner, N. N. Choi, M.-H. Lee, A. Czasch, and R. D\"{o}ner, J. Phys. B 40, F157 (2007).
\bibitem{Arseth1974} S. J. Arseth and K. Zare, Celest. Mech. 10, 185 (1974).
\bibitem{Choi2004} N. N. Choi, M.-H. Lee, and G. Tanner, Phys. Rev. Lett. 93, 054302 (2004).
\bibitem{Lee2005} M.-H. Lee, G. tanner, and N. N. Choi, Phys. Rev. E 71, 056208 (2005).
\bibitem{Hao1998} B.-L. Hao and W.-M. Zheng, {\it Applied Symbolic Dynamics and Chaos} (World Scientific, Singapore 1998).
\bibitem{Bai1998} Z.-Q. Bai, Y. Gu, and J.-M. Yuan, Physica D 118, 17 (1998).\bibitem{Lee2005b} M.-H. Lee, N. N. Choi, and G. Tanner, Phys. Rev. E 72, 066215 (2005).
\bibitem{Lee2010JKPS} M.-H. Lee and N. N. Choi, J. Kor. Phys. Soc. 56, 1799 (2010).
\bibitem{Du1988} M. L. Du and J. B. Delos, Phys. Rev. A 38, 1896; ibid. 38, 1913 (1988).
\bibitem{Tomsovic1994} S. Tomsovic and D. Ullmo, Phys. Rev. E 50, 145 (1994).
\end{references}
\end{document}